\documentclass[aps,pra,twocolumn,superscriptaddress,nofootinbib,longbibliography,groupedaddress]{revtex4-1}

\usepackage{amsmath}
\usepackage{amscd}
\usepackage{wasysym}
\usepackage[hidelinks]{hyperref}
\usepackage{color}
\usepackage{braket}
\usepackage{float}

\usepackage[T1]{fontenc}
\usepackage{ae,aecompl}
\usepackage{amsfonts}
\usepackage{amsthm}
\usepackage{dsfont}
\usepackage{graphicx}
\usepackage{amsfonts}
\usepackage{physics}

\newcommand{\ue}{\mathrm{e}}

\newcommand{\la}{\langle}
\newcommand{\ra}{\rangle}

\newcommand{\figref}[1]{Fig.~\ref{#1}}
\newcommand{\secref}[1]{Sec.~\ref{#1}}

\begin{document}
	
	
	\title{Impossibility of bosonic autonomous entanglement engines in the weak-coupling limit}

	
	\author{Bradley Longstaff}
	\author{Michael G. Jabbour}
	\author{Jonatan Bohr Brask}
	\affiliation{Center for Macroscopic Quantum States (bigQ), Department of Physics,
		Technical University of Denmark, 2800 Kongens Lyngby, Denmark}

	
	\begin{abstract}
		Entanglement is a fundamental feature of quantum physics and a key resource for quantum communication, computing and sensing. Entangled states are fragile and maintaining coherence is a central challenge in quantum information processing. Nevertheless, entanglement can be generated and stabilised through dissipative processes. In fact, entanglement has been shown to exist in the steady state of certain interacting quantum systems subject solely to incoherent coupling to thermal baths. This has been demonstrated in a range of bi- and multipartite settings using systems of finite dimension. Here we focus on the steady state of infinite-dimensional bosonic systems. Specifically, we consider any set of bosonic modes undergoing excitation-number-preserving interactions of arbitrary strength and divided between an arbitrary number of parties that each couple weakly to thermal baths at different temperatures. We show that a unique steady state is always separable.
	\end{abstract}
	
	\maketitle

	\section{Introduction}
	\label{sec.intro}
	
	Entanglement is a central feature of quantum physics and a key resource for many quantum-information-processing tasks \cite{Horodecki2009}, such as quantum communication, computing and sensing ~\cite{Gisin07,giovanetti2011,Pironio2016,Pirandola20}. The generation and stabilisation of entangled states is a challenging task. As entanglement is generally fragile with respect to loss and noise~\cite{Horodecki2009}, considerable effort is usually required to isolate the system of interest from the environment. Nevertheless, it has been shown that coupling to an environment can be used to aid entanglement generation~\cite{GenEnt1,GenEnt2,GenEnt3,GenEnt4,GenEnt5,GenEnt6,GenEnt7,Naseem22}. Steady-state entanglement can also be generated by dissipation and external driving~\cite{SteadyState1,SteadyState2,SteadyState3,SteadyState4}. In fact, dissipative steady-state entanglement generation is even possible without any source of coherence or external control, a setting referred to as autonomous. This has been demonstrated in a number of different contexts for finite-dimensional systems~\cite{NoExtForce1,NoExtForce2,NoExtForce3,NoExtForce4,NoExtForce5,NoExtForce6,NoExtForce7,Brask2015,Tavakoli2018,Man2019,Tavakoli2020,Brask2022}.
	
	Much less is known for infinite-dimensional systems, although they are widely employed in practice in experiments with quantum optical, optomechanical, and superconducting setups. In particular, the class of so-called Gaussian states and operations in bosonic systems are ubiquitous as they are relatively simple to realise and their theory is well understood ~\cite{Weedbrook2012}. Gaussian processes, e.g. squeezing and linear interferometers, are used extensively in continuous-variable quantum information processing and quantum optics in general. However, it turns out that non-Gaussian resources are necessary in order to perform various important quantum-information-processing tasks including, for example, entanglement distillation~\cite{CVdistillation1,CVdistillation2,CVdistillation3}, quantum error correction~\cite{CVerrorCorr}, and universal quantum computation~\cite{CVcomp1,CVcomp2}. It is therefore of interest to determine the combinations of Gaussian and non-Gaussian resources that can generate steady-state entanglement. In the present work, we demonstrate that autonomous, steady-state entanglement in bosonic systems with passive Gaussian interactions is not possible.
	
	We consider a multi-mode bosonic system undergoing quadratic, excitation-preserving interactions (i.e., passive Gaussian interactions or linear interferometers) of arbitrary strength, divided between an arbitrary number of parties. The modes of different parties couple weakly to thermal baths at different temperatures, with the energy distribution in each bath following either Bose-Einstein or Fermi-Dirac statistics (both bosonic and spin baths are allowed). The system-bath interactions are also quadratic but not necessarily excitation number preserving. We show that in such a setting, no entanglement can ever be present in the steady state between the parties of the system.
	
	The paper is organised as follows. In Section~\ref{sec:ThermalMachine} we introduce the model of our thermal machines. We then present a separability criterion in Section~\ref{sec:GaussEntanglement} that will be satisfied by the steady states. The master equations that describe the dynamics of our machines are derived in Section~\ref{sec:MEandSS}. We focus on two situations: a global approach, which is appropriate when the interactions between subsystems is strong relative to the system-bath couplings; and a local approximation, which describes a situation where the interactions between subsystems is weak relative to the system-bath couplings. We then show that the steady state is separable in both situations. Finally, we conclude in Section~\ref{sec:Conc}.

	\section{Bosonic autonomous thermal machine}
	\label{sec:ThermalMachine}
	
\begin{figure*}[t]
		\centering
		\includegraphics[width=0.73\textwidth]{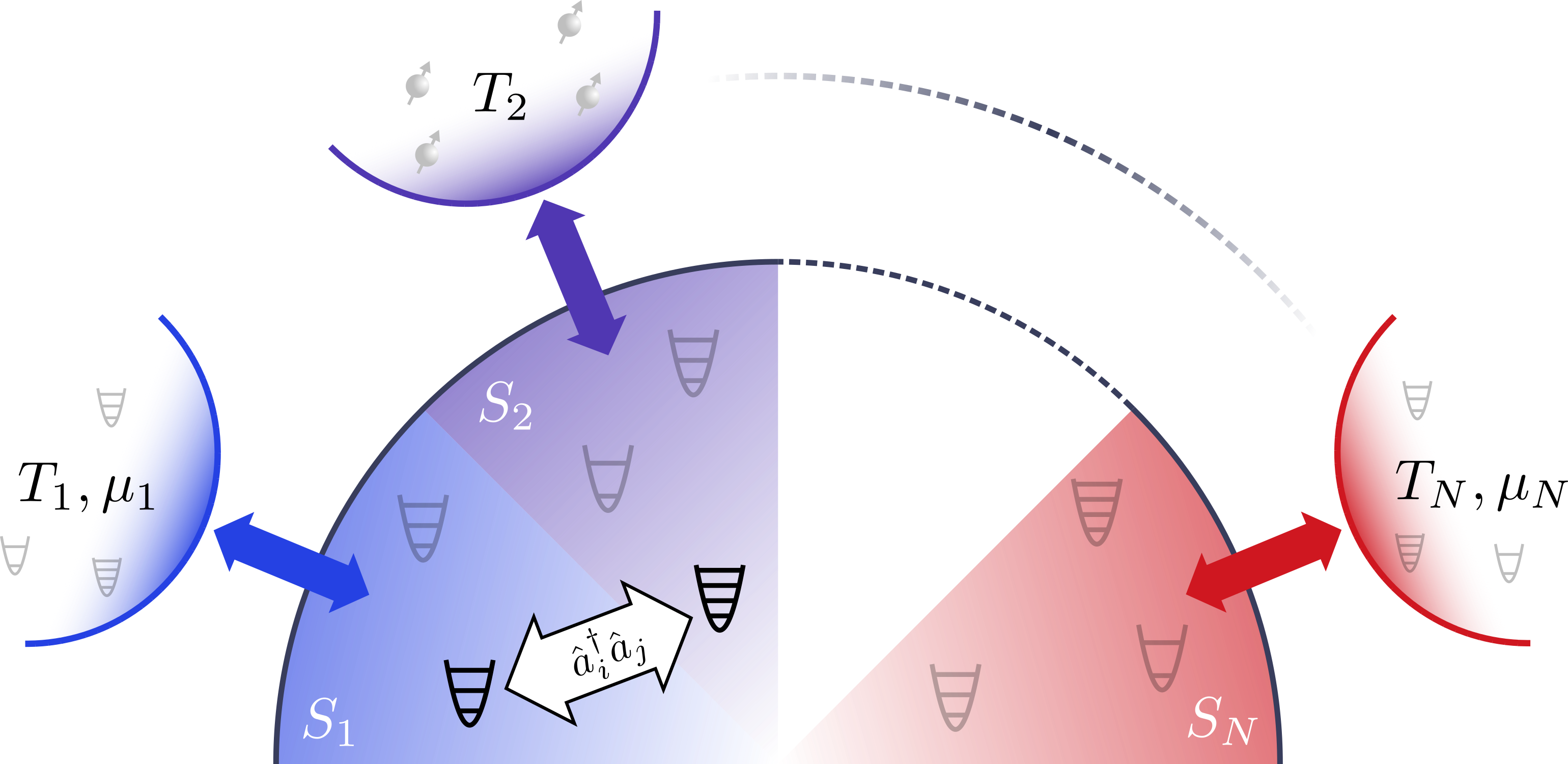}
		\caption{Sketch of the quantum thermal machine. The system consists of $d$ bosonic modes distributed between $N$ subsystems. The system modes are coupled via excitation-number-preserving interactions. The subsystems are separately coupled to either bosonic or spin baths. Bosonic baths exchange both energy and particles with the system while the spin baths exchange only energy. The system-bath coupling is taken to be weak, while the inter-system couplings can be of arbitrary strength.}\label{fig:Machine}
	\end{figure*}

	We consider a collection of bosonic modes undergoing excitation-preserving interactions of arbitrary strength and distributed between an arbitrary number of parties that each couple weakly to thermal baths at different temperatures (\figref{fig:Machine}).
	The total system $S$ consists of $d$ bosonic modes with associated creation and annihilation operators that satisfy the canonical commutation relations $[\hat{a}_j,\hat{a}^\dag_k] = \delta_{jk}$, $[\hat{a}_j,\hat{a}_k] = 0$ and $[\hat{a}^\dag_j,\hat{a}^\dag_k] = 0$. The corresponding quadrature operators for each mode are defined as
	\begin{equation}
		\hat{q}_j = \frac{1}{\sqrt{2}}\left(\hat{a}^\dag_j + \hat{a}_j\right), \quad \hat{p}_j = \frac{i}{\sqrt{2}}\left(\hat{a}^\dag_j - \hat{a}_j\right),
	\end{equation}
	and fulfil the commutation relations $[\hat{q}_j,\hat{p}_k] = i\delta_{jk}$, where we use units with $\hbar = 1$ throughout.
	
	The Hamiltonian of the system $\hat{H}_S$ is taken to be a general stable ($\hat{H}_S > 0$), quadratic, and excitation-number-preserving Hamiltonian
		\begin{equation}
			\label{eqn:ham s}
			\hat{H}_S = \sum_{i,j} \mathcal{H}_{ij}\hat{a}^\dag_i \hat{a}_j ,
		\end{equation} 
		where $\mathcal{H}_{ij}$ is a Hermitian matrix. There are $N$ subsystems and baths, with the $n$\textsuperscript{th} subsystem denoted by $S_n$ and coupling to the $n$\textsuperscript{th} bath $E_n$. The total environment is described by the continuous non-interacting Hamiltonian
	\begin{equation}
		\hat{H}_E = \sum_{n=1}^{N} \hat{H}_{E_n} = \sum_{n=1}^{N} \int \epsilon_{n}(q) \, \hat{c}^\dag_{n}(q) \hat{c}_{n}(q)\,dq,
	\end{equation}
	and we assume that the bath spectra are non negative with $\epsilon_{n}(q)\geq 0$. In general, each bath may be bosonic or consist of spins. For the bosonic baths, the bath creation and annihilation operators satisfy the canonical commutation relations
	\begin{align}
		[\hat{c}_{n}(k),\hat{c}^\dag_{m}(q)] &= \delta_{nm}\delta (k-q),\\
		[\hat{c}_{n}(k),\hat{c}_{m}(q)] &= [\hat{c}^\dag_{n}(k),\hat{c}^\dag_{m}(q)] = 0,
	\end{align}
	while spins in different spin baths commute and spins within the same bath obey the anticommutation relations
	\begin{align}
		\{\hat{c}_{n}(k),\hat{c}^\dag_{n}(q)\} &= \delta (k-q),\\
		\{\hat{c}_{n}(k),\hat{c}_{n}(q)\} &= \{\hat{c}^\dag_{n}(k),\hat{c}^\dag_{n}(q)\} = 0.
	\end{align}
	Finally, the system-bath interaction is a quadratic  Hamiltonian of the form $\hat{H}_{SE} = \sum_{n=1}^N \hat{V}_n$, with 
	\begin{equation}
		\hat{V}_n = \sum_{j \in S_n} \int g_{j}(q)\left(\hat{a}^\dag_j + \eta_j \hat{a}_j\right)\hat{c}_n(q) \, dq + h.c..
	\end{equation}
	Here the $j$-sum is taken over the modes of subsystem $S_n$, the complex coefficients $g_j(q)$ quantify the strength of each interaction, and the parameters $\eta_j \in \{0,1\}$ allow for both excitation-number-conserving ($\eta_j = 0$) and `position-like' ($\eta_j = 1$) coupling to the environment .
	
	The bosonic baths exchange energy and particles with the system, and are thus characterised by an inverse temperature $\beta_n = 1/k_BT_n$ and chemical potential $\mu_n$. On the other hand, the spin baths only exchange energy with the system. We assume that the environment is initially in the thermal state
	\begin{equation}
		\label{eqn:rhoE}
		\hat{\rho}_E = \bigotimes_{n=1}^{N} \frac{\ue^{-(\hat{H}_{E_n}-\mu_n \hat{N}_n)\beta_n}}{\tr\left[\ue^{-(\hat{H}_{E_n}-\mu_n \hat{N}_n)\beta_n}\right]},
	\end{equation}
	where $\hat{N}_n = \int \hat{c}^\dag_{n}(q)\hat{c}_{n}(q)\, dq$ is the number operator of the $n$\textsuperscript{th} bath and the chemical potential should only be included for bosonic baths. In this case, the two-point correlation functions satisfy 
	\begin{align}
		\la \hat{c}^\dag_{n}(k)\hat{c}_{m}(q)\ra = p_n(\epsilon_{n}(k)) \, \delta_{nm} \, \delta (k-q),\\
		\la \hat{c}_{n}(k)\hat{c}_{m}(q)\ra = 0, \quad \la \hat{c}^\dag_{n}(k)\hat{c}^\dag_{m}(q)\ra = 0,	
	\end{align}
	where $\la \cdot \ra$ is the expectation value in the state $\hat{\rho}_E$ and
	\begin{equation}
		p_n(\epsilon) = \left[\xi_n + \ue^{(\epsilon-\mu_n)\beta_n}\right]^{-1}
	\end{equation}
	is the Fermi-Dirac distribution for spin baths ($\xi_n = 1$) or the Bose-Einstein distribution for bosonic baths ($\xi_n = -1$).

	\section{A criterion for separability}
	\label{sec:GaussEntanglement}
	
	Our goal is to derive the steady state of the system in the model above. Before proceeding, however, we first consider a criterion for separability. As the steady state is found to be Gaussian (see Section \ref{sec:SSs}), separability can be determined from the covariance matrix. By definition, a Gaussian state $\hat{\rho}$ has a Wigner function of the form
	\begin{equation}
		W(x) = \frac{1}{(2\pi)^d}\frac{1}{\sqrt{\det \Sigma}}\ue^{-\frac{1}{2}(x-\bar{x})^T\Sigma^{-1}(x-\bar{x})}.
	\end{equation}
	Here $x = (q_1,p_1,\ldots,q_d,p_d) \in \mathbb{R}^{2d}$ are canonical phase-space coordinates, $\bar{x}$ is a vector of first moments $\bar{x}_j = \tr[\hat{x}_j\hat{\rho}]$ with $\hat{x} = (\hat{q}_1,\hat{p}_1,\ldots,\hat{q}_d,\hat{p}_d)$, and the covariance matrix $\Sigma$ is a real, symmetric and positive-definite matrix with elements
	\begin{equation}
		\Sigma_{jk} = \frac{1}{2}\la \hat{x}_j\hat{x}_k + \hat{x}_j \hat{x}_k \ra - \la \hat{x}_j \ra\la \hat{x}_k \ra ,
	\end{equation}
	where $\la \cdot \ra = \tr[\cdot \hat{\rho}]$. The covariance matrix of a valid quantum state satisfies the Robertson-Schr\"odinger uncertainty relation
	\begin{equation}
		\label{eq:physicalCov}
		\Sigma + \frac{i}{2} \Omega \geq 0 ,
	\end{equation}
	where $\Omega$ is the symplectic matrix
	\begin{equation}
		\Omega = \bigoplus_{j=1}^d \begin{pmatrix} 0 & 1 \\ -1 & 0 \end{pmatrix}.
	\end{equation}
	
	As shown by Werner and Wolf \cite{Werner2001}, a bipartite Gaussian state of systems $A$ and $B$ with covariance matrix $\Sigma_{AB}$ is separable if and only if there exist valid covariance matrices $\Sigma_A$ and $\Sigma_B$ for systems $A$ and $B$ such that $\Sigma_{AB} \geq \Sigma_A \oplus \Sigma_B$. We would like to apply this result to our steady states. However, it turns out that the steady-state covariance matrices derived below in \secref{sec:MEandSS} are more conveniently expressed in coordinates different to $(q_1,p_1,\ldots,q_d,p_d)$. Specifically, in the complex coordinates $(q_1+i p_1,\ldots,q_d+i p_d,q_1-i p_1,\ldots,q_d-i p_d)/\sqrt{2}$ they have the simple block structure
	\begin{equation}
		\label{eqn:sig steady}
		\Sigma' = \begin{pmatrix} \sigma^\ast & 0 \\ 0 & \sigma \end{pmatrix} ,
	\end{equation}
	where $\sigma_{jk} = \frac{1}{2} \la \hat{a}^\dag_j \hat{a}_k + \hat{a}_k \hat{a}^\dag_j\ra - \la \hat{a}^\dag_j\ra\la\hat{a}_k\ra$ and the prime denotes the complex form. The transformation to complex coordinates is achieved by first applying the permutation $P$ that maps $(q_1,p_1,\ldots,q_d,p_d)$ to $ (q_1,\ldots,q_d,p_1,\ldots,p_d)$, and then the unitary matrix
	\begin{equation}
		U = \frac{1}{\sqrt{2}}\begin{pmatrix} I_d & iI_d \\ I_d & -iI_d \end{pmatrix},
	\end{equation}
	where $I_d$ is the $d \times d$ identity matrix. Thus, the similarity transformation $UP$ yields $\Sigma' = UP\Sigma P^\dag U^\dag$. 
	
	From the steady-state covariance matrix \eqref{eqn:sig steady}, we observe that $\la \hat{a}_j \hat{a}_k \ra - \la \hat{a}_j \ra\la \hat{a}_k \ra = 0$, and so the corresponding Gaussian steady state is not squeezed. Separability of the steady state then follows from the fact that every entangled Gaussian state is squeezed, see e.g. \cite{Wolf03}. For completeness, we provide a proof that any $\Sigma'$ of the form \eqref{eqn:sig steady} corresponds to a $\Sigma$ that satisfies the separability condition across any bipartition. In complex coordinates, the physicality condition \eqref{eq:physicalCov} becomes $\Sigma' + \frac{1}{2} K \geq 0$, where
	\begin{equation}
		K = \begin{pmatrix} I_d & 0 \\0 & -I_d \end{pmatrix} ,
	\end{equation}
	as may be seen by applying the transformation $UP$. Since $\Sigma'$ in \eqref{eqn:sig steady} comes from a valid steady state, it follows that $\sigma - \frac{1}{2} I_d \geq 0$. Taking the complex conjugate, we also have $\sigma^\ast - \frac{1}{2} I_d \geq 0$. These two results then imply that $\Sigma' \geq \frac{1}{2} I_{2d}$. Transforming back to the real coordinates we obtain $\Sigma \geq \frac{1}{2}I_{2d}$, as the eigenvalues are preserved under the transformation. For any bipartition of the total system we have that $I_{2d} = I_A \oplus I_B$, with $I_A$ and $I_B$ identity matrices corresponding to subsystems A and B, and $\frac{1}{2}I$ is always a valid covariance matrix (the covariance matrix of the vacuum state). Hence, a Gaussian state with covariance matrix of the form~\eqref{eqn:sig steady} is separable across every bipartition.

	\section{Open-system dynamics and steady states}
	\label{sec:MEandSS}
	
	In this section we derive master equations for the open-system dynamics of the setup in \figref{fig:Machine}. It is shown that the steady state is a Gaussian state with a covariance matrix of the form \eqref{eqn:sig steady}, and is therefore separable by the argument of \secref{sec:GaussEntanglement}. We consider weak system-bath couplings and large baths, and assume that the initial composite state of the system and baths factorises. In this case, one can apply a perturbative approach in the system-bath coupling to derive a master equation for the time evolution of the system. A time-local (Markovian) description of the system evolution, which does not depend on the entire history of the system state, is obtained provided that correlations in the baths decay much faster than variations of the state of the system $\tilde{\rho}(t)$ in the interaction picture (denoted by a tilde).
	
	Following the standard procedure, one finds (see, e.g., Refs.~\cite{Rivas12,Potts2019} for an in-depth derivation) 
	\begin{equation} \label{eqn:rho dot 0}
		\begin{aligned}
			& \frac{d}{dt}\tilde{\rho}(t) \\
			& = -\int_0^\infty du \, \textnormal{tr}_E\bigg(\left[\hat{H}_{SE}(t),\left[\hat{H}_{SE}(t-u),\tilde{\rho}(t)\otimes\hat{\rho}_E\right]\right]\bigg).
		\end{aligned}
	\end{equation}
	Substituting in the expressions for $\hat{H}_{SE}$ and $\hat{\rho}_E$ leads to
	\begin{widetext}
		\begin{equation} \label{eqn:rho dot 1}
			\begin{aligned}
				& \frac{d}{dt}\tilde{\rho}(t)  = \sum_{n=1}^{N} \sum_{j,k\in S_n} \int_0^\infty du \, \bigg( \tr_E\left[\hat{B}^\dag_{jn}(u)\hat{B}_{kn}\hat{\rho}_E\right] \left(\hat{A}_j^\dag(t-u)\tilde{\rho}(t)\hat{A}_k(t) - \hat{A}_k(t)\hat{A}_j^\dag(t-u)\tilde{\rho}(t)\right) \\
				& \qquad \qquad \qquad \qquad \qquad \qquad \quad + \tr_E\left[\hat{B}_{jn}(u)\hat{B}^\dag_{kn}\hat{\rho}_E\right] \left(\hat{A}_j(t-u)\tilde{\rho}(t)\hat{A}_k^\dag(t)-\hat{A}_k^\dag(t)\hat{A}_j(t-u)\tilde{\rho}(t)\right) + h.c. \bigg),
			\end{aligned}
		\end{equation}
	\end{widetext}
	%
	%
	where we have introduced the system operators $\hat{A}_j(t) = \hat{a}_j(t) + \eta_j \hat{a}^\dag_j(t)$ to simplify the equation, and in the interaction picture the bath operators $\hat{B}_{jn} = \int g^\ast_{j}(q)\hat{c}_{n}(q)\,dq$ have the form
	\begin{equation}
		\hat{B}_{jn}(u) = \int g^\ast_{j}(q)\hat{c}_{n}(q)\ue^{-i\epsilon_{n}(q)u}\, dq.
	\end{equation}
	It is also convenient to define the one-sided Fourier transforms of the bath correlation functions,
	\begin{align}
		C^{(n,1)}_{jk}(\omega) &= \int_0^\infty \tr_E\left[\hat{B}^\dag_{jn}(u)\hat{B}_{kn}\hat{\rho}_E\right]\ue^{-i\omega u}du,\label{eqn:C coeff1}\\
		C^{(n,2)}_{jk}(\omega) &= \int_0^\infty \tr_E\left[\hat{B}_{jn}(u)\hat{B}^\dag_{kn}\hat{\rho}_E\right] \ue^{i\omega u} du,
	\end{align}
	and split them into their Hermitian and anti-Hermitian parts
	\begin{equation}
		\label{eqn:herm aherm}
		C^{(n,l)}_{jk}(\omega) = \frac{1}{2} \gamma^{(n,l)}_{jk}(\omega)+ i s^{(n,l)}_{jk}(\omega).
	\end{equation}
	The following expressions will also be useful when we consider separability of the steady state,
	\begin{align}
		\tr_E\left[\hat{B}^\dag_{jn}(u)\hat{B}_{kn}\hat{\rho}_E\right] &= \int g_{j}(q)g^\ast_{k}(q)\ue^{i\epsilon_{n}(q)u}p_n(\epsilon_{n}(q))\,dq , \label{eqn:bath corr 1} \\
		\tr_E\left[\hat{B}_{jn}(u)\hat{B}^\dag_{kn}\hat{\rho}_E\right] & = \int g^\ast_{j}(q)g_{k}(q)\ue^{-i\epsilon_{n}(q)u} \nonumber \\
		& \qquad \qquad \times [1-\xi_n p_n(\epsilon_{n}(q))]\,dq. \label{eqn:useful2}
	\end{align}

	In order to make further progress with the evolution equation \eqref{eqn:rho dot 1}, we need to deal with the $u$ dependence appearing in the interaction-picture creation and annihilation operators. We consider two situations separately and derive a master equation for each case. First we take the global approach, in which the baths interact with the eigenmodes of the total system Hamiltonian. This is appropriate when the interaction between subsystems is strong relative to the system-bath interactions and results in a so-called global master equation \cite{Hofer2017}. Second, we make a local approximation, where each bath interacts locally with the eigenmodes of the subsystem to which it couples. This is appropriate when the interaction between subsystems is weaker than the system-bath couplings (or comparable, see \cite{Hofer2017}), and results in a local master equation. Note that the validity regimes of the local and global master equations have some overlap \cite{Hofer2017}, and hence treating both enables us to apply our separability result in the entire range from very weak to strong inter-system interactions. In both cases we apply the full secular approximation to obtain a physical (completely positive trace-preserving) master equation. The condition for this approximation to be valid differs between the local and global approaches, and therefore the details are provided separately below.
	
	\subsection{Global master equation}
	\label{sec:Secular}
	
	Let us first derive a global master equation. To this end, we follow~\cite{Abbruzzo21} and make use of the Bogoliubov transformation. The system Hamiltonian \eqref{eqn:ham s} can be written in the form $\hat{H}_S = \mathbf{\hat{a}^\dag}\mathcal{H}\mathbf{\hat{a}}$, where $\mathbf{\hat{a}^\dag} = (\hat{a}^\dag_1,\ldots,\hat{a}^\dag_d)$ and $\mathcal{H}$ is a Hermitian matrix that can be diagonalised with a unitary matrix $U$. The transformation $\mathbf{\hat{a}} = U \mathbf{\hat{b}}$ then brings the system Hamiltonian into the form
		\begin{equation}
			\hat{H}_S = \sum_j \omega_j \hat{b}^\dag_j \hat{b}_j,
		\end{equation}
		where $\omega_j > 0$ are the eigenvalues of $\mathcal{H}$, and the new creation and annihilation operators $\hat{b}_j,\hat{b}^\dag_j$ satisfy the usual bosonic canonical commutation relations.
	
	Starting from \eqref{eqn:rho dot 1} and applying a full-secular approximation yields a Markovian master equation for the system state $\hat{\rho}(t)$ (in the Schr\"odinger picture) that is completely positive and trace preserving,
	%
	\begin{align}
		\label{eqn:full sec me}
		&\frac{d}{dt}\hat{\rho}(t) = -i[\hat{H}_S+\hat{H}_{LS},\hat{\rho}(t)] + \nonumber \\ &\sum_{\lambda=1}^D \sum_{u,v\in \Lambda_\lambda} \left[\Gamma_{uv}^{(1)}(\omega_\lambda)\mathcal{D}(\hat{b}_u^\dag,\hat{b}_v^\dag) + \Gamma_{uv}^{(2)}(\omega_\lambda)\mathcal{D}(\hat{b}_u,\hat{b}_v)\right][\hat{\rho}(t)].
	\end{align}
	%
	%
	Here we have introduced the superoperator $\mathcal{D}(\hat{X}_i,\hat{X}_j)$ that, given the two operators $\hat{X}_i,\hat{X}_j$, acts on a quantum state $\hat{\rho}$ as
	\begin{equation}
		\mathcal{D}(\hat{X}_i,\hat{X}_j)[\hat{\rho}] = \hat{X}_i\hat{\rho}\hat{X}_j^\dag - \frac{1}{2}\{\hat{X}_j^\dag \hat{X}_i,\hat{\rho}\}.
	\end{equation}
	The system has $D$ different energy eigenspaces labelled by $\lambda$, where $\Lambda_\lambda$ is the set of normal modes associated with the $\lambda$\textsuperscript{th} eigenspace corresponding to eigenvalue $\omega_\lambda$, and
	\begin{align}
		\Gamma^{(1)}_{uv}(\omega) &= \sum_{n=1}^{N}\sum_{i,j\in S_n}\gamma^{(n,1)}_{ij}(\omega)U^\ast_{iu} U_{jv},\label{eqn:gam1 full}\\
		\Gamma^{(2)}_{uv}(\omega) &= \sum_{n=1}^{N}\sum_{i,j\in S_n}\gamma^{(n,2)}_{ij}(\omega)U_{iu} U^\ast_{jv},\label{eqn:gam2 full}
	\end{align}
	where the $\gamma^{(n,k)}_{ij}(\omega)$ are defined in \eqref{eqn:herm aherm}. The unitary part of the dynamics is generated by the system Hamiltonian
	\begin{equation}
		\hat{H}_S = \sum_{\lambda=1}^D \sum_{u\in \Lambda_\lambda} \omega_\lambda \hat{b}^\dag_u \hat{b}_u
	\end{equation}
	and the so-called Lamb-shift Hamiltonian
	\begin{equation}
		\hat{H}_{LS} = \sum_{\lambda=1}^D \sum_{u,v\in\Lambda_\lambda} \varphi_{uv}(\omega_\lambda)\hat{b}^\dag_u \hat{b}_v,
	\end{equation}
	where
	\begin{equation}
		\begin{aligned}
			\varphi_{uv}(\omega) & = \sum_{n=1}^{N} \sum_{i,j\in S_n}\left[\left(s_{ij}^{(n,1)}(\omega)+s_{ji}^{(n,2)}(\omega)\right) \right. \\
			& + \left. \left(s^{(n,1)}_{ji}(-\omega)+s^{(n,2)}_{ij}(-\omega)\right)\eta_i \eta_j\right]U^\ast_{iu}U_{jv},
		\end{aligned}
	\end{equation}
	and the $s^{(n,k)}_{ij}(\omega)$ are also defined in \eqref{eqn:herm aherm}. The Lamb-shift Hamiltonian commutes with $\hat{H}_S$ and so produces a shift in the energy levels of $\hat{H}_S$.
	
	The full-secular approximation, which was required to arrive at the master equation \eqref{eqn:full sec me}, is valid provided that the differences $\nu$ of the normal-mode eigenvalues satisfy $\min_{\nu\neq\nu'}|\nu-\nu'|>1/\tau_B$, where $\tau_B$ is the largest correlation time of any of the baths. Furthermore, this approximation has removed all terms from the master equation that could generate squeezing, i.e., only passive terms remain.
	
	\subsection{Local master equation}
	\label{sec:Local}
	
	Now suppose that the $N$ subsystems are weakly interacting with each other. In this case the local approach may be justified, where the coupling between the subsystems is neglected when calculating the effects of the environment. We thus begin by writing the system Hamiltonian in the form $\hat{H}_S = \sum_{n}\hat{H}_n + \hat{H}_{C}$, where $\hat{H}_n$ is the local Hamiltonian for subsystem $S_n$ and $\hat{H}_{C}$ accounts for the coupling between the different subsystems. Each $\hat{H}_n$ is excitation-number-conserving and, as before, can be put into diagonal form
	\begin{equation}
		\hat{H}_n = \sum_{j=1}^{M_n}\omega_{nj}\hat{b}_{nj}^\dag \hat{b}_{nj}
	\end{equation}
	with a Bogoliubov transformation
	\begin{equation}
		\hat{a}_i = \sum_{j=1}^{M_n} U^n_{ij}\hat{b}_{nj}.
	\end{equation}
	Here $M_n$ is the number of modes in subsystem $S_n$, $\omega_{nj}>0$ are the eigenvalues of $\mathcal{H}_n$, $\hat{b}_{nj}$ are the Bogoliubov operators associated with subsystem $S_n$, and $i \in S_n$. The components $U_{ij}^n$ can be viewed as the elements of a unitary matrix $U^n$ after a relabelling of the $i$ index.
	
	These transformations do not diagonalise $\hat{H}_S$ and will typically result in a complicated coupling term $\hat{H}_{C}$. However, this does not complicate the derivation, as the coupling between the subsystems is ignored when calculating the influence of the environment. With this in mind, starting from \eqref{eqn:rho dot 1} and applying a full-secular approximation yields a Markovian master equation that is again completely positive and trace preserving
	%
	\begin{align} 
		\label{eqn:local me}
		&\frac{d}{dt}\hat{\rho}(t) = -i[\hat{H}_S+\hat{H}_{LS},\hat{\rho}(t)] + \nonumber \\ &\sum_{n=1}^N\sum_{\lambda=1}^{D_n} \sum_{u,v\in n_\lambda} \bigg[\Gamma_{uv}^{(n,1)}(\omega_{n\lambda})\mathcal{D}(\hat{b}_{nu}^\dag,\hat{b}_{nv}^\dag) + \nonumber \\ & \hspace{8em} \Gamma_{uv}^{(n,2)}(\omega_{n\lambda})\mathcal{D}(\hat{b}_{nu},\hat{b}_{nv})\bigg][\hat{\rho}(t)].
	\end{align}
	%
	Here $D_n$ is the number of energy eigenspaces associated with the subsystem $S_n$, and $n_\lambda$ is the set of normal modes associated with the $\lambda$\textsuperscript{th} eigenspace corresponding to the eigenvalue $\omega_{n\lambda}$. Furthermore,
	\begin{align}
		\Gamma^{(n,1)}_{uv}(\omega) &= \sum_{i,j\in S_n}\gamma^{(n,1)}_{ij}(\omega){U^n_{iu}}^\ast U^n_{jv},\\
		\Gamma^{(n,2)}_{uv}(\omega) &= \sum_{i,j\in S_n}\gamma^{(n,2)}_{ij}(\omega)U^n_{iu} {U^n_{jv}}^\ast,
	\end{align}
	where the $\gamma^{(n,k)}_{ij}(\omega)$ are defined in \eqref{eqn:herm aherm}. 
	
	While the mode transformations simplify each local subsystem Hamiltonian $\hat{H}_n$, their effect on the coupling Hamiltonian $\hat{H}_C$ leads to a system Hamiltonian of the form
	\begin{equation}
		\label{eqn:system ham local approach}
		\hat{H}_S = \sum_{n=1}^N\sum_{\lambda=1}^{D_n} \sum_{u\in n_\lambda}\omega_{n \lambda}\hat{b}^\dag_{n u}\hat{b}_{n u} + \hat{H}_C'.
	\end{equation}
	The contribution $\hat{H}_C'$ contains terms of the form $\hat{b}_{nu}^\dag\hat{b}_{mv}$, which involve Bogoliubov operators for different subsystems and energy eigenspaces. However, the full details of the coefficients appearing in this part are not needed in what follows. The coupling of the system to the baths produces an additional contribution to the unitary part of the dynamics
	\begin{equation}
		\label{eqn:local lamb}
		\hat{H}_{LS} = \sum_{n=1}^N\sum_{\lambda = 1}^{D_n}\sum_{u,v\in n_\lambda} \varphi^{(n)}_{uv}(\omega_{n \lambda}) \hat{b}^\dag_{n u}\hat{b}_{n v},
	\end{equation}
	where
	\begin{equation}
		\begin{aligned}
			\varphi^{(n)}_{uv}(\omega) & = \sum_{i,j\in S_n} \left[\left(s_{ij}^{(n,1)}(\omega)+s_{ji}^{(n,2)}(\omega)\right) \right. \\
			& + \left. \left(s^{(n,1)}_{ji}(-\omega)+s^{(n,2)}_{ij}(-\omega)\right)\eta_i \eta_j \right] {U^n_{iu}}^\ast U^n_{jv},
		\end{aligned}
	\end{equation}
	and the $s^{(n,k)}_{ij}(\omega)$ are also defined in \eqref{eqn:herm aherm}. In contrast to the global approach, this term does not commute with the system Hamiltonian and therefore does not simply shift the energies of the system Hamiltonian.
	
	In the local approach, the validity of the full-secular approximation is determined by the local Hamiltonians of each subsystem. For each subsystem, the differences $\nu_n$ of the normal-mode eigenvalues associated with the local Hamiltonian $\hat{H}_n$ must satisfy $\min_{\nu_n\neq\nu_n'}|\nu_n-\nu_n'|>1/\tau_B$. Again, this approximation results in a master equation containing only passive terms that cannot generate squeezing.
	
	\subsection{The steady-state solution}
	\label{sec:SSs}
	
	We now show that whenever the steady state of the global or local master equation is unique, it is a Gaussian state with a covariance matrix of the form \eqref{eqn:sig steady}. Thus, for any initial system state the steady state will be a separable Gaussian state. We first provide a short proof that assumes the steady state is unique, and then derive a condition for the steady state to be unique. Both derivations make use of the fact that the global and local master equations take Gaussian states to Gaussian states. This is because the Hamiltonian part is quadratic in the Bogoliubov operators, and the Lindblad operators in the dissipative part are linear in the Bogoliubov operators.
	
	For the simple proof, note that the right hand side of both the global \eqref{eqn:full sec me} and local \eqref{eqn:local me} master equations can be written in terms of a Liouvillian $\mathcal{L}$ as $\mathcal{L}\hat{\rho}(t)$. In each case $\mathcal{L}$ commutes with the superoperator $\mathcal{N}\boldsymbol{\cdot} = \left[\hat{N},\boldsymbol{\cdot}\right]$, where the total excitation-number operator is $\hat{N} = \sum_{j} \hat{b}_j^\dag \hat{b}_j$ in the global approach and $\hat{N} = \sum_{n,j} \hat{b}_{nj}^\dag \hat{b}_{nj}$ in the local approach. Therefore, any state that is initially block diagonal in the $\hat{N}$ eigenbasis, with no coherences between different number sectors, will remain block diagonal for all time. Such a state has no squeezing and thus a covariance matrix of the separable form \eqref{eqn:sig steady}. This proves our result provided that the steady state is unique because an initial Gaussian state with covariance matrix of the form \eqref{eqn:sig steady} must then remain a Gaussian state with a covariance matrix of this form for all time. 
	
	We now derive a condition for the steady state to be unique. Recall that any initial system state $\hat{\rho}(0)$ can be written as
	\begin{equation}
		\label{eqn:cs basis}
		\hat{\rho}(0) = \frac{1}{\pi^{2d}}\int \la {\alpha}|\hat{\rho}(0)|{\beta}\ra|{\alpha}\ra\la{\beta}| d^2{\alpha}d^2{\beta},
	\end{equation}
	where $|{\alpha}\ra$ is a $d$-mode Glauber coherent state with amplitudes ${\alpha}=(\alpha_1,\ldots,\alpha_d)$. By the linearity of the master equations, if we obtain the long-time evolution of each term $|{\alpha}\ra\la{\beta}|$ in the decomposition \eqref{eqn:cs basis} we can then deduce the form of the steady state. 
	
	Let us first deal with the coherent states $|{\alpha}\ra\la {\alpha}|$ appearing in \eqref{eqn:cs basis}. These are Gaussian states, each with the same covariance matrix $I_{2d}/2$ but different centres. We now show that in the long-time limit every coherent state $|{\alpha}\ra\la {\alpha}|$ tends to the same Gaussian state $\hat{\rho}_G$ with zero first moments and a covariance matrix of the form \eqref{eqn:sig steady}. Because the master equations map Gaussian states to Gaussian states, the initial coherent states $|{\alpha}\ra$ will remain Gaussian for all time. It is then enough to work out the steady-state expectation values of the first and second moments.
	
	We focus on the global master equation \eqref{eqn:full sec me}, as the derivation for the local master equation proceeds along the same lines. From \eqref{eqn:full sec me} it follows that the time evolution of an operator $\hat{O}$ is determined by
	\begin{widetext}
		\begin{equation}
			\label{eqn:heis full sec}
			\frac{d}{dt}\hat{O} = i[\hat{H}_S+\hat{H}_{LS},\hat{O}] + \sum_{\lambda=1}^D \sum_{u,v\in \Lambda_\lambda} \Gamma_{uv}^{(1)}(\omega_\lambda)\left(\hat{b}_v \hat{O}\hat{b}^\dag_u -\frac{1}{2}\{\hat{b}_v\hat{b}^\dag_u,\hat{O}\}\right)
			+\Gamma_{uv}^{(2)}(\omega_\lambda)\left(\hat{b}^\dag_v \hat{O}\hat{b}_u -\frac{1}{2}\{\hat{b}^\dag_v\hat{b}_u,\hat{O}\}\right).
		\end{equation}
	\end{widetext}
	The equations of motion for the one and two-point correlators $\la \hat{b}_i\ra$ and $\la \hat{b}_i \hat{b}_j \ra$ are thus obtained by setting $\hat{O}$ to $\hat{b}_i$ and $\hat{b}_i \hat{b}_j$ respectively. Now recall that every coherent state $|{\alpha}\ra$ has the same covariance matrix, $I_{2d}/2$, and that the Bogoliubov operators are related to the original operators by the unitary transformation $\mathbf{\hat{a}} = U \mathbf{\hat{b}}$. From this we see that $\la \hat{b}_i \hat{b}_j \ra - \la\hat{b}_i\ra\la\hat{b}_j\ra = 0$ for all $i,j$ at time $t=0$. By plugging these initial conditions into the equations of motion for $\beta_{ij} = \la \hat{b}_i\hat{b}_j\ra - \la\hat{b}_i\ra\la\hat{b}_j\ra$ it is straightforward to check that $\beta_{ij} = 0$ for all time (see Appendix \ref{app:bbeta}). Therefore, $\la \hat{a}_i \hat{a}_j \ra  -\la \hat{a}_i\ra\la\hat{a}_j\ra= 0$ at every time for all $i,j$. Because the equations of motion for the (co)variances do not couple to the first moments, the covariance matrix of every initial coherent state $|{\alpha}\ra$ tends to the same steady-state covariance matrix of separable form \eqref{eqn:sig steady}.
	
	All that remains is to show that the centre of every initial coherent state goes to the same value in the long-time limit. For this we need the equations of motion for the first moments
	\begin{align}
		\label{eqn:first moments eom}
		\frac{d}{dt}\la\hat{b}_j\ra &= -i\omega_{j}\la \hat{b}_j \ra - i\sum_{u \in \Lambda_j}\varphi_{ju}(\omega_{j})\la \hat{b}_u\ra \nonumber \\ &\qquad+\frac{1}{2}\sum_{u \in \Lambda_j}\left(\Gamma^{(1)}_{ju}(\omega_{j})-\Gamma^{(2)}_{uj}(\omega_{j})\right)\la \hat{b}_u \ra.
	\end{align}
	%
	For simplicity we have redefined $\omega_j$ as the eigenvalue associated with the mode $\hat{b}_j$ and $\Lambda_j$ as the eigenspace to which the eigenvalue $\omega_j$ belongs. This set of linear equations can be written in the form $ d {b} /dt = R {b}$, where $b_i = \la \hat{b}_i \ra$ and $R$ is a non-Hermitian matrix. The Hermitian part of $R$ is negative semidefinite (see Appendix~\ref{app:HermitianRneg}), and therefore, if $R$ is invertible, then ${b} = {0}$ is the only steady-state solution. If this condition is satisfied, it follows that every initial state $|{\alpha}\ra$ tends to the same Gaussian state $\hat{\rho}_G$ with zero first moments and a covariance matrix of the form \eqref{eqn:sig steady}.
	
	The long-time dynamics of the cross terms $|{\alpha}\ra\la{\beta}|$ (${\alpha} \neq {\beta}$) appearing in \eqref{eqn:cs basis} require a little more work. The Wigner-Weyl transformation of $|{\alpha}\ra\la{\beta}|$ is a complex Gaussian in phase space that is highly oscillatory. A familiar example where terms like this appear is in the Wigner function of a cat state. For Lindblad master equations with linear Lindblad operators and quadratic Hamiltonians, the coupling to the environment exponentially damps the oscillatory parts of these phase-space functions. This is a manifestation of decoherence. Therefore, the complex oscillatory parts cannot persist in the steady state. The dynamics of complex Gaussian wave packets generated by equations of Lindblad type were analysed in \cite{Graefe18}. Using these results, we find that each $|{\alpha}\ra\la{\beta}|$ tends to $N_{\alpha\beta}\hat{\rho}_G$. Here $\hat{\rho}_G$ is the same Gaussian steady state we found before, and the $N_{\alpha\beta}$ are normalisation factors that guarantee the steady state $\hat{\rho}_\infty$ is normalised. We can then conclude that the steady state $\hat{\rho}_\infty$ is the separable Gaussian state $\hat{\rho}_G$. 
	
	The same procedure can be followed with the local master equation \eqref{eqn:local me} to obtain the same result. The only difference, barring additional indices, is the coupling term $\hat{H}_C'$ appearing in the system Hamiltonian \eqref{eqn:system ham local approach}. However, this does not pose any complications when following the steps above, as $\hat{H}_C'$ only includes terms of the form $\hat{b}_{nu}^\dag\hat{b}_{mv}$ and cannot introduce any squeezing.

	\section{Conclusion \label{sec:Conc}}
	
	We have shown that autonomous, steady-state entanglement generation is impossible in bosonic systems of arbitrary size undergoing quadratic excitation-number-preserving interactions (i.e., passive Gaussian interactions or linear interferometers) and weakly coupled to thermal bosonic and/or spin baths at different temperatures. Our result applies to both the local and global regimes, which cover the entire range from weak to strong inter-system coupling. Our result holds provided that: the system Hamiltonian is stable ($\hat{H}_S>0$); a secular approximation is justified; and the steady state is unique. This result contrasts with previous findings for finite-dimensional autonomous thermal machines with excitation-preserving interactions, which can generate steady-state entanglement strong enough to exhibit both steering and nonlocality. Passive Gaussian interactions, even combined with non-bosonic baths, are sufficiently restrictive to preclude any steady-state entanglement. We note that the condition for the steady state to be unique ($R$ invertible) will most likely be satisfied in all practical situations, where the model parameter values will never be completely symmetric.
	
	Having ruled out steady-state entanglement in this case, it is interesting to ask whether bosonic steady-state entanglement can be generated in more complicated settings. One natural next step would be to include higher-order inter-system interactions, such as three- or four-wave mixing. The challenge here may be to determine whether the (generally non-Gaussian) steady-state is separable or not. One could also attempt to go beyond the weak-coupling limit by applying, for example, the reaction coordinate method \cite{Nazir2018}.
	
	\begin{acknowledgements}
		We thank Patrick Potts for enlightening discussions and feedback. This work was supported
		by a research grant (40864) from VILLUM
		FONDEN, as well as by the Carlsberg Foundation CF19-0313 and the Independent Research Fund Denmark 7027-00044B.
	\end{acknowledgements}

	\bibliography{bengine}
	
	\appendix
	\section{Time evolution of the covariance matrix elements $\beta_{jk}$}\label{app:bbeta}
		The equations of motion for the covariance matrix elements $\beta_{jk} = \langle \hat{b}_j \hat{b}_k\rangle -\langle \hat{b}_j\rangle \langle \hat{b}_k\rangle$ can be obtained from the operator evolution equation \eqref{eqn:heis full sec}. We first compute the equations for $\langle \hat{b}_j\rangle$ and $\langle \hat{b}_j \hat{b}_k\rangle$, and then combine the results to find
		\begin{align}
			\label{eqn:beta time evo}
			\frac{d}{dt}\beta_{jk} = &-i\left(\omega_{j}+\omega_{k}\right)\beta_{jk} +\nonumber\\ &\sum_{u\in\Lambda_j}f_{ju}(\omega_j)\beta_{ku} + \sum_{u\in\Lambda_k}f_{ku}(\omega_k)\beta_{ju}.
		\end{align}
		For simplicity we have redefined $\omega_j$ as the eigenvalue associated with mode $\hat{b}_j$ and $\Lambda_j$ as the eigenspace to which $\omega_j$ belongs. The functions $f_{iu}(\omega)$ are defined as
		\begin{equation}
			f_{iu}(\omega) = \frac{1}{2}\left[\Gamma_{iu}^{(1)}(\omega)-\Gamma_{ui}^{(2)}(\omega)\right] - i \varphi_{iu}(\omega),
		\end{equation}
		in terms of functions appearing in the main text. The key result of this Appendix is that when each element $\beta_{jk} = 0$ at $t=0$, then each $\beta_{jk}$ is zero for all time. This follows immediately from \eqref{eqn:beta time evo}.

	\section{The Hermitian part of \texorpdfstring{$R$}{R} is negative semidefinite \label{app:HermitianRneg}}
	
	Let the Hermitian part of $R$ be denoted by $R_H = (R+R^\dag)/2$. From the equations of motion for the first moments in the main text \eqref{eqn:first moments eom} we obtain the matrix elements
	\begin{equation}
		[R_H]_{ij} = \frac{1}{2}\left(\Gamma^{(1)}_{ij}(\omega_i)-\Gamma^{(2)}_{ji}(\omega_i)\right)\delta_{\omega_i,\omega_j},
	\end{equation}
	where $\omega_i$ is the eigenvalue associated with the Bogoliubov mode $\hat{b}_i$. We permute the rows and columns of $R_H$ to bring it into block diagonal form, where each block corresponds to the same eigenvalue. After a relabelling of indices, the block $B$ corresponding to the eigenvalue $\omega$ has the elements
	\begin{equation}
		\label{eqn:elements apx}
		B_{ij} = \frac{1}{2}\left(\Gamma^{(1)}_{ij}(\omega)-\Gamma^{(2)}_{ji}(\omega)\right).
	\end{equation}
	We now show that this block is negative semidefinite.
	
	Making use of the expressions \eqref{eqn:C coeff1}-\eqref{eqn:useful2} and \eqref{eqn:gam1 full}-\eqref{eqn:gam2 full} we find that the matrix elements in \eqref{eqn:elements apx} can be written as
	\begin{equation}
		\begin{aligned}
			B_{ij} & = \pi\sum_{n=1}^N \int \tilde{g}_{in}(q)\tilde{g}^\ast_{jn}(q)\delta(\epsilon_n(q)-\omega) \\
			& \qquad \qquad \quad \times \left((1+\xi_n)p_n(\epsilon_n(q))-1\right) \, dq,
		\end{aligned}
	\end{equation}
	where we have defined
	\begin{equation}
		\tilde{g}_{in}(q) = \sum_{j\in S_n} U^\ast_{ji}g_j(q)
	\end{equation}
	in terms of the unitary matrix $U$ that diagonalises the system Hamiltonian and the coupling parameters between the system and environment $g_j(q)$. For every complex vector $x$
	\begin{equation}
		\label{eqn:negative def}
		\begin{aligned}
			x^\dag B x & = \pi\sum_{n=1}^N \int |x_n(q)|^2 \delta(\epsilon_n(q)-\omega) \\
			& \qquad \qquad \quad \times \left((1+\xi_n)p_n(\epsilon_n(q))-1\right)\,dq,
		\end{aligned}
	\end{equation}
	with $x_n(q) = \sum_j x_j \tilde{g}^\ast_{jn}(q)$. For the bosonic baths, $\xi_n = -1$ and the contribution to \eqref{eqn:negative def} is less than or equal to zero. For the spin baths, $\xi_n = +1$ but $p_n(\omega)<1/2$ since $\omega > 0$. The contribution from the spin baths is therefore also less than or equal to zero, and it follows that $x^\dag B x \leq 0$. This holds for every block of $R_H$, and so the Hermitian part of $R$ is negative semidefinite.

\end{document}